\documentclass[prb,showpacs,superscriptaddress,preprintnumbers,amssymb,twocolumn,amsfonts,nofootinbib]{revtex4}
\usepackage{graphicx}
\usepackage{amssymb}
\usepackage{amsmath}
\usepackage{bm}

\begin{document}

\input epsf
\def\beq{\begin{equation}}
\def\eeq{\end{equation}}
\def\beqn{\begin{eqnarray}}
\def\eeqn{\end{eqnarray}}
\def\etal{\emph{et al.}}
\renewcommand{\thefootnote}{\fnsymbol{footnote}}

\title{Topological insulators beyond the Brillouin zone via Chern parity}

\author{Andrew M.~Essin}
\affiliation{Department of Physics, University of California,
Berkeley, CA 94720}
\author{J.~E.~Moore}
\affiliation{Department of Physics, University of California,
Berkeley, CA 94720} \affiliation{Materials Sciences Division,
Lawrence Berkeley National Laboratory, Berkeley, CA 94720}
\date{\today}

\begin{abstract}
The topological insulator is an electronic phase stabilized by spin-orbit coupling that supports propagating edge states and is not adiabatically connected to the ordinary insulator.  In several ways it is a spin-orbit-induced analogue in time-reversal-invariant systems of the integer quantum Hall effect (IQHE).  This paper studies the topological insulator phase in disordered two-dimensional systems, using a model graphene Hamiltonian introduced by Kane and Mele as an example.  The nonperturbative definition of a topological insulator given here is distinct from previous efforts in that it involves boundary phase twists that couple only to charge, does not refer to edge states, and can be measured by pumping cycles of ordinary charge.  In this definition, the phase of a Slater determinant of electronic states is determined by a Chern parity analogous to Chern number in the IQHE case.  Numerically we find, in agreement with recent network model studies, that the direct transition between ordinary and topological insulators that occurs in band structures is a consequence of the perfect crystalline lattice.  Generically these two phases are separated by a metallic phase, which is allowed in two dimensions when spin-orbit coupling is present.  The same approach can be used to study three-dimensional topological insulators.
\end{abstract}
\maketitle

\section{Introduction}

Considerable theoretical and experimental effort has been devoted to the quest for an intrinsic spin Hall effect~\cite{murakami,sinova,kato,wunderlich} that would allow generation of spin currents by an applied electric field.  Interesting mechanisms for such spin current generation make use of spin-orbit coupling, which breaks the $SU(2)$ spin symmetry of free electrons but not time-reversal symmetry.  A dissipationless type of intrinsic spin Hall effect was predicted~\cite{kane&mele1-2005, kane&mele2-2005} to arise in materials that have an electronic energy gap. This ``quantum spin Hall effect'' (QSHE) in certain materials with time-reversal symmetry has a subtle relationship to the integer quantum Hall effect, in which time-reversal symmetry is explicitly broken by a magnetic field.

In a system with unbroken time-reversal symmetry, a dissipationless charge current is forbidden, but a dissipationless transverse spin current is allowed, of the form
\beq
{\cal J}^i_j = \alpha \epsilon_{ijk} E_k.
\label{spintranseq}
\eeq
The current on the left is a spin current and $\epsilon$ is the fully antisymmetric tensor.  Note that a spin current requires two indices, one for the direction of the current and one for the direction of angular momentum that is transported.  The constant of proportionality $\alpha$ depends on the specific mechanism: for example, the (dissipative) extrinsic D'yakonov-Perel mechanism~\cite{dyakonovperel} predicts a small $\alpha$ that depends on impurity concentration.  The QSHE builds on the construction by Haldane~\cite{haldane-1988} of a lattice  ``Chern insulator'' model, with broken time-reversal symmetry but without net magnetic flux, that shows a $\nu = 1$ IQHE.  The simplest example of a QSHE is obtained by taking two copies of Haldane's model, one for spin-up electrons along some axis and one for spin-down.  Time-reversal symmetry can be maintained if the effective IQHE magnetic fields are opposite for the two spin components.  Then an applied electric field generates a transverse current in one direction for spin-up electrons, and in the opposite direction for spin-down electrons.  There is no net charge current, consistent with time-reversal symmetry, but there is a net spin current.  However, models like this in which one component of spin is perfectly conserved are both unphysical, since realistic spin-orbit coupling does not conserve any component, and not very novel, since for each spin component the physics is exactly the same as Haldane's model and the spin components do not mix.

\begin{figure}[!ht]
	\scalebox{0.4}{\includegraphics{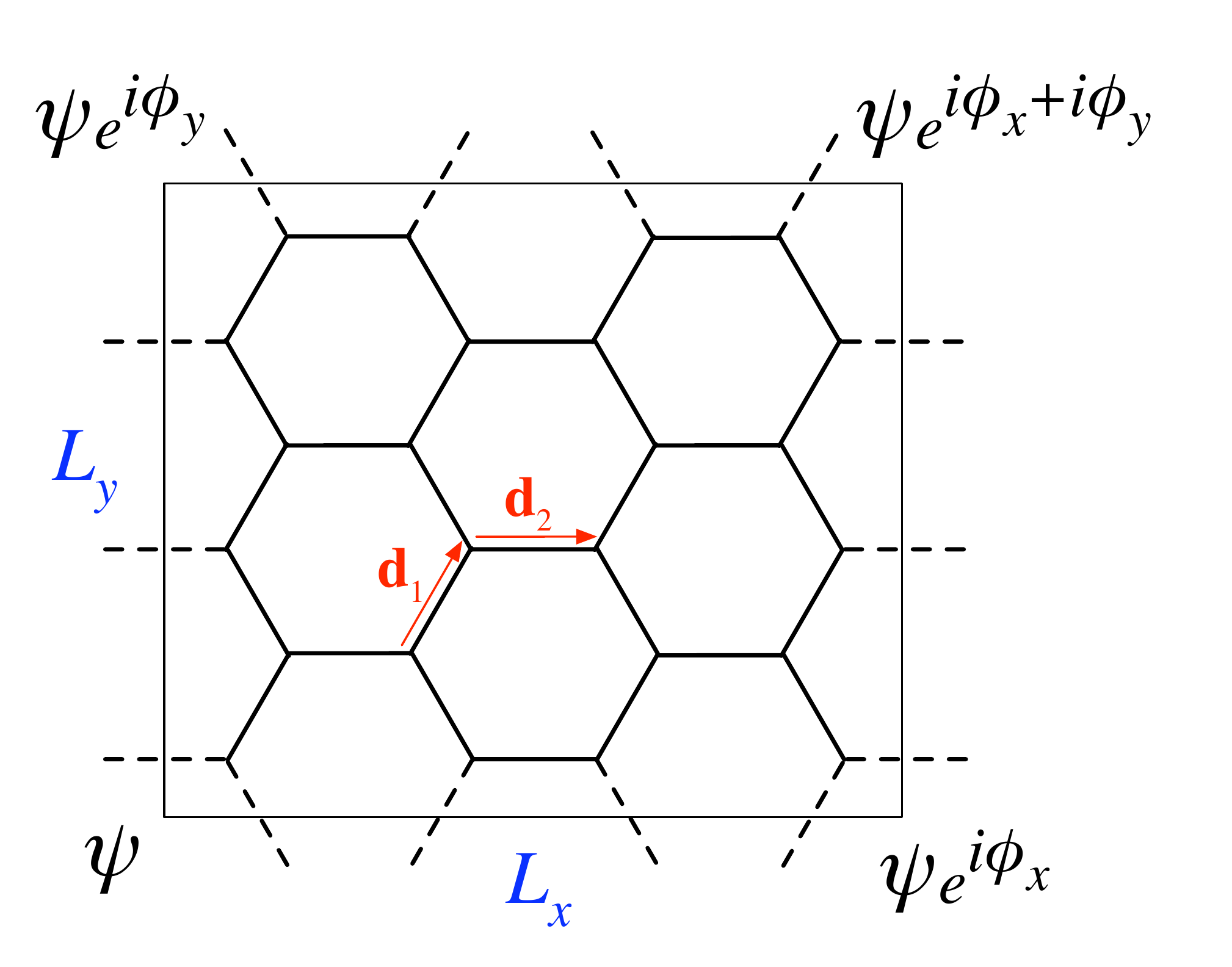}}
	\caption{The honeycomb lattice on which the tight-binding Hamiltonian resides.
		For the two sites depicted, the factor $\nu_{ij}$ of equation (\ref{HSO}) 
		is $\nu_{ij} = -1$.  The phases $\phi_{x,y}$ describe twisted boundary
		conditions, introduced in equation \eqref{bdyconds}.
	\label{honeycomb}}
\end{figure}

More subtle physics emerges when one asks how the QSHE appears in more realistic band structures.  Remarkably, band insulators of noninteracting two-dimensional electrons with spin-orbit coupling divide into two classes: the ``ordinary insulator'', which in general has no propagating edge modes and no spin Hall effect, and the ``topological insulator'', which has stable propagating edge modes and a generic spin Hall effect, although the amount of spin transported (the coefficient $\alpha$ in (\ref{spintranseq})) is nonuniversal.  These phases are associated with a $\mathbb{Z}_2$-valued topological invariant (an ``oddness'' or ``evenness'', in the language of parity\footnote{Throughout ``parity'' is used to denote oddness or evenness, rather than a spatial inversion eigenvalue.}) in the same way that IQHE phases are associated with an integer-valued topological invariant.  For explicitness, consider the model of graphene introduced by Kane and Mele.\cite{kane&mele1-2005}  This is a tight-binding model for independent electrons on the honeycomb lattice  (Fig.~\ref{honeycomb}).  The spin-independent part of the Hamiltonian consists of a nearest-neighbor hopping, which alone would give a semimetallic spectrum with Dirac nodes at certain points in the 2D Brillouin zone, plus a staggered sublattice potential whose effect is to introduce a gap:
\beq\label{H0}
H_0 = t \sum_{\langle i j \rangle \sigma} c^\dagger_{i \sigma} c_{j \sigma} + \lambda_v \sum_{i \sigma} \xi_i c^\dagger_{i \sigma} c_{i \sigma}.
\eeq
Here $\langle ij \rangle$ denotes nearest-neighbor pairs of sites, $\sigma$ is a spin index, $\xi_i$ alternates sign between sublattices of the honeycomb, and $t$ and $\lambda_v$ are parameters.

The insulator created by increasing $\lambda_v$ is an unremarkable band insulator.  However, the symmetries of graphene also permit an ``intrinsic'' spin-orbit coupling of the form
\beq\label{HSO}
H_{SO} = i \lambda_{SO} \sum_{\langle \langle ij \rangle \rangle \sigma_1 \sigma_2} \nu_{ij} c^\dagger_{i \sigma_1} s^z_{\sigma_1 \sigma_2} c_{j \sigma_2}.
\eeq
Here $\nu_{ij} = (2 / \sqrt{3}) \hat{\bm{d}}_1 \times \hat{\bm{d}}_2 = \pm 1$, where $i$ and $j$ are next-nearest-neighbors and $\hat{\bm{d}}_1$ and $\hat{\bm{d}}_2$ are unit vectors along the two bonds that connect $i$ to $j$.  The Hamiltonian $H_0+H_{SO}$ conserves $s^z$, the distinguished component of electron spin, and reduces for fixed spin (up or down) to Haldane's model.\cite{haldane-1988}

The ``topological insulator'' phase created when $|\lambda_{SO} | \gg | \lambda_v| $ is quite different from the ordinary insulator that appears when $|\lambda_v| \gg |\lambda_{SO}|$ (here we assume that there is an energy gap between the lower and upper band pairs in which the  Fermi level lies).  The former has counterpropagating edge modes and shows the QSHE, while the latter does not.\cite{kane&mele1-2005}  Does this phase exist for more realistic spin-orbit coupling?  The spin component $s_z$ is no longer a good quantum number when the Rashba spin-orbit coupling is added:
\beq\label{HR}
H_R = i \lambda_R \sum_{\langle ij \rangle \sigma_1 \sigma_2} c^\dagger_{i \sigma_1} \left (\bm{s}_{\sigma_1 \sigma_2} \times \bm{\hat d}_{ij}\right)_z c_{j \sigma_2},
\eeq
with $\bm{d}_{ij}$ the vector from $i \rightarrow j$ and $\bm{\hat d}_{ij}$ the corresponding unit vector.  (Note that Rashba spin-orbit coupling is not intrinsic to graphene but generated by inversion-symmetry breaking in the out-of-plane direction.)  The topological insulator survives but is strongly modified in the presence of this term.  For a general 2D band structure with $s^z$ conserved, there are many phases labeled by an integer $n$, as in the IQHE: if spin-up electrons are in the $\nu=n$ state, then spin-down electrons must be in the $\nu=-n$ state by time-reversal symmetry, where the sign indicates that the direction of the effective magnetic field is reversed.  Once $s^z$ is not conserved, there are only two insulating phases, the ``ordinary'' and ``topological'' insulators.  A heuristic definition of the topological insulator, without reference to any particular spin component or the spin Hall effect, is as a band insulator that is required to have gapless propagating edge modes at the sample boundaries.  The decoupled $\nu = \pm n$ cases with $s^z$ conserved are adiabatically connected, once $s^z$ is not conserved, to the ordinary insulator for even $n$ and to the topological insulator for odd $n$.  A review of how these two cases emerge as the only possibilities in 2D follows in Section II.

It is not obvious at first glance how to generalize the topological insulator phase to finite, noncrystalline systems, rather than band structures, as when the parameters of the Hamiltonian $H=H_0 + H_{SO}+ H_R$ are drawn from a random distribution.  The first approach was in terms of a spin Chern number~\cite{shenghaldane} similar to the Chern integer in finite IQHE systems, but there is now agreement that for a clean band structure the only invariants are of $\mathbb{Z}_2$ type, rather than integer type.\cite{fu&kane1-2006,moore&balents-2006,fukui&hatsugai-2006}  Two equivalent definitions of the appropriate $\mathbb{Z}_2$ invariant for a finite disordered system, in the simple case when the disorder splits all degeneracies other than Kramers degeneracies, are as follows (the full definition is given and compared to previous work in the following section).  The finite system can be considered as a unit ``supercell'' of a large 2D lattice.  A large, finite supercell gives many bands, but each pair of bands connected by time reversal (Kramers pair) can be assigned its own $\mathbb{Z}_2$ invariant.\cite{moore&balents-2006}   The phase of the supercell system, if the Fermi level lies in a gap, is then identified by adding up all the invariants (mod 2).  Alternately, a direct definition of the phase in the finite system can be given that is related to the notion of ``$\mathbb{Z}_2$ pumping''.\cite{fu&kane1-2006}  Real charge is pumped as the flux through the system is taken from $0$ to $hc/2 e$ ({\it half} the usual flux quantum that appears in IQHE pumping); we show that in the topological insulator, any pumping cycle, properly defined, pumps an odd number of electron charges, while for the ordinary insulator any cycle pumps an even number of charges.

We implement this definition numerically using an explicit algorithm introduced by Fukui and Hatsugai~\cite{fukuihatsugai} for computing $\mathbb{Z}_2$ topological invariants on a Brillouin zone.  The topological insulator phase is robust to disorder: while different realizations of disorder assign different ``Chern parities'' to individual subbands, it is found that the total for occupied subbands is always ``odd'' for a wide range of parameters, which in our definition indicates a topological insulator.  In the IQHE, a pair of bands of opposite Chern number can annihilate as the strength of disorder is increased; in the QSHE, two band pairs that both have odd Chern parity can annihilate, i.e., become two even-parity band pairs.  If the topological insulator can be destroyed by band annihilation, then there are extended (i.e., topologically nontrivial) states with an arbitrarily small gap; it may be the case that for some range of parameters, there are extended states at the Fermi level even in the thermodynamic limit, indicating a metallic phase.  In the IQHE, there is only a single energy with extended states rather than a range of energies, and hence no metallic phase.  We find the phase diagram of the graphene model with on-site disorder, and in the presence of non-zero Rashba coupling find evidence for a metallic phase intervening between ordinary and topological insulators.

Recent work by Obuse \etal~\cite{obuse&alii-2007} obtains a phase diagram and critical exponents using a network model for the spin quantum Hall effect that is similar to the Chalker-Coddington network model\cite{chalker&coddington-1988} for the IQHE (see also Onoda \etal\cite{onoda&alii-2006} for a quasi-1D study of localization in the Kane-Mele Hamiltonian with disorder).  Our results on the phase diagram are consistent with these works, although our method is unable to generate large enough system sizes to confirm the exponents found for the phase transitions.  To understand how the network and Chern-parity approaches complement each other, consider the integer quantum Hall effect (IQHE): while the phenomenological network approach to the IQHE is valuable both to find the critical indices precisely and to identify the minimal necessary elements of a theory for the transition, Chern-number studies remain important for studies of effects such as the floating of extended states,\cite{laughlinfloating,khmelnitskiifloating,yang&bhatt-1996} where knowledge of the topological properties of a state is required.  The network model gives more accurate information about the phase transitions but, if only the localization length is probed, does not distinguish the different phases in bulk.  A more technical difference between the two approaches is discussed at the end of Section II.

Section II reviews how the topological insulator phase in perfect crystals arises from a parity-valued topological invariant of the band structure, similar to the TKNN~\cite{thouless&alii-1982} integers in the IQHE.  It then gives two mathematically equivalent definitions of the topological insulator phase in disordered systems based on Chern parity.  One definition simply considers a finite disordered system as a supercell of an infinite lattice system, while the other is based upon closed charge pumping cycles driven by application of flux to a finite periodic system.  Section III reviews the Fukui-Hatsugai algorithm~\cite{fukuihatsugai} adapted to numerical computation of these invariants in disordered systems, then computes the phase diagram of the Kane-Mele graphene model~\cite{kane&mele1-2005} with on-site disorder.  The conclusions of our study for general 2D disordered systems are summarized in a short Section IV.  While there is a three-dimensional version of the QSHE~\cite{moore&balents-2006,rroy3d,fu&alii-2007,bernevigchen} that is less directly connected to the IQHE and has interesting localization behavior, we will restrict our attention to 2D except for some comments in the final section.

\section{Chern parities for disordered noninteracting electron systems}

\subsection{The definition of the $\mathbb{Z}_2$ invariant in clean systems}

We review one definition of the $\mathbb{Z}_2$ invariant of a band pair in a 2D band structure, then explain its generalization to noncrystalline systems.
With the Hamiltonian 
\begin{equation}\label{ham}
	H = H_0 + H_{SO} + H_R
\end{equation}
defined in equations \eqref{H0}, \eqref{HSO}, and \eqref{HR}, or any periodic, single-electron Hamiltonian, a Berry connection $\bm{\mathcal{A}}$ can be defined on the Brillouin zone (BZ) from the periodic part $u(\bm{k})$ of a Bloch state $\psi_{\bf k} = u({\bf k}) e^{i {\bf k} \cdot {\bf r}}$.  For a single nondegenerate band, we define $\bm{\mathcal{A}}_{j}(\bm{k}) = -i\langle u_{j}| \bm{\nabla}\!_{\bm{k}} |u_{j} \rangle$ for band $j$.  This Berry connection can be used to understand how a commensurate magnetic field in a two-dimensional lattice system, which allows definition of Bloch states on the (magnetic) Brillouin zone, leads to the IQHE.  The TKNN integers~\cite{thouless&alii-1982} in the IQHE can be written as integrals over the BZ of the Berry field strength $\mathcal{F} = (\bm{\nabla}\!_{\bm{k}}\times\bm{\mathcal{A}})_z$,
\beqn
n_j &=& -\frac{1}{2\pi} \int_{BZ}\,\mathcal{F}_j\,d^2 k \cr
&=& \frac{i}{2 \pi} \int_{BZ}\, \left[ \left\langle \frac{\partial u}{\partial k_x} \Big| \frac{\partial u}{\partial k_y} \right\rangle - \left \langle \frac{\partial u}{\partial k_y} \Big| \frac{\partial u}{\partial k_x} \right\rangle \right].
\eeqn

The TKNN integers are integer-valued topological invariants: the TKNN integer of a band cannot change as long as the band remains nondegenerate.  When bands touch, only the total TKNN integer of the bands is well-defined,~\cite{avron&alii-1983} and the IQHE state of a gapped system is given by the sum of TKNN integers for occupied bands.  
However, all these integers vanish in a time-reversal-invariant band structure (see, e.g., Ref.~\onlinecite{moore&balents-2006}).  Instead there is a $\mathbb{Z}_2$ invariant associated with a band {\it pair} in a time-reversal-invariant 2D Fermi system.\cite{kane&mele2-2005}  
Time-reversal invariance requires that
\begin{equation}\label{trev}
	\Theta \mathcal{H}(-\bm{k}) \Theta^{-1} = \mathcal{H}(\bm{k}),
\end{equation}
where $\mathcal{H}(\bm{k})$ is the Bloch Hamiltonian and $\Theta = -i\sigma_y K$ is the action of time reversal ($K$ performs complex conjugation and $\sigma_y$, the usual Pauli matrix, acts on spin indices).

With time-reversal breaking, as in the IQHE, it is reasonable to ignore band degeneracies, but time-reversal invariance forces the single electron energies to be doubly degenerate at certain time-reversal-invariant momenta (see, for example, Ref.~\onlinecite{sakurai4.4}).  These degeneracies are known as ``Kramers degeneracies''.  The single-band connection $\bm{\mathcal{A}}_j$ will not be globally defined.  So long as each Kramers pair remains separated by gaps in the energy spectrum from all others, the appropriate generalization is
\begin{align}\label{A}
	\bm{\mathcal{A}}_j &= -i\left(
			\langle u_{j1}| \bm{\nabla}\!_{\bm{k}} |u_{j1}\rangle +
			\langle u_{j2}| \bm{\nabla}\!_{\bm{k}} |u_{j2}\rangle
			\right) \notag \\
		&= -i\, \mathrm{tr}\, \bm{u}_{j}\!^\dagger \bm{\nabla}\!_{\bm{k}} \bm{u}_{j}.
\end{align}
In the second equation, $\bm{u}_{j} = \left(u_{j1},u_{j2} \right)$ where the Bloch functions are viewed as column vectors, i.e., $\bm{u}_{j}$ is a matrix.  This compact notation follows Fukui and Hatsugai\cite{fukui&hatsugai-2006} and makes it simple to include more than one band pair simply by adding more columns to $\bm{u}$.  Finally, we again define a field strength associated with the potential, $\mathcal{F} = (\bm{\nabla}\!_{\bm{k}}\times\bm{\mathcal{A}})_z$.

\begin{figure}[!ht]
	\centering
	\scalebox{0.35}{\includegraphics{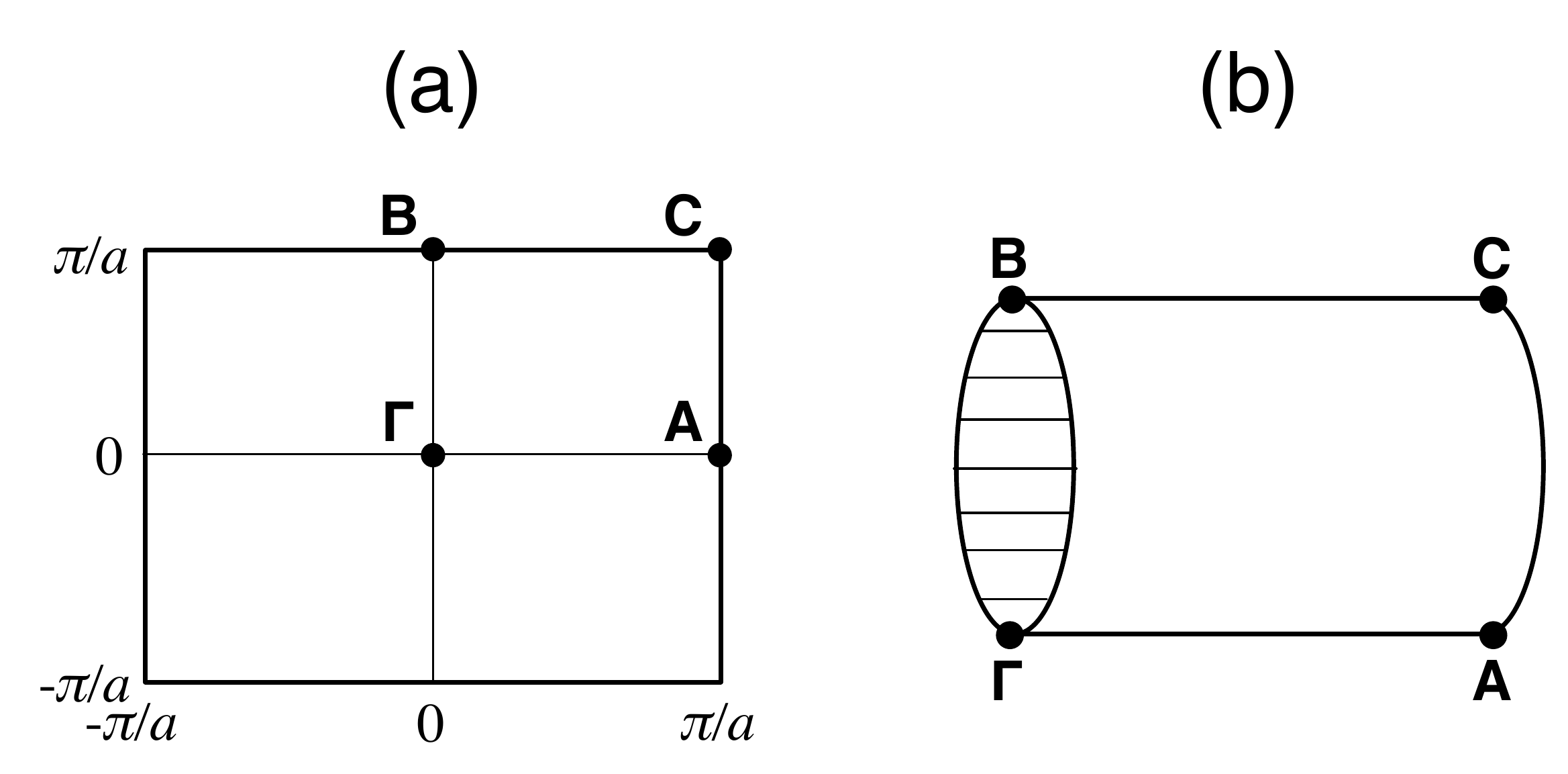}}
	\caption{(a) A two-dimensional Brillouin zone; note that any such Brillouin
	  zone, including that for graphene, can be smoothly deformed to a torus.  The
	  labeled points are time-reversal-invariant momenta.
		(b) The effective Brillouin zone (EBZ).  The horizontal lines on the boundary
		circles $\partial(\text{EBZ})$ connect time-reversal-conjugate points, where
		the Hamiltonians are related by time reversal and so cannot be specified
		independently.
		\label{figebz}}
\end{figure}

From these quantities, Fu and Kane\cite{fu&kane1-2006} give the following formula for the $\mathbb{Z}_2$ topological invariant in terms of the Bloch functions of the clean system:
\begin{equation}\label{Dk}
	D = \frac{1}{2\pi} \left[ \oint_{\partial(EBZ)} d\bm{k}\cdot\!\bm{\mathcal{A}}
			- \int_{EBZ} d^2\!k\,\mathcal{F} \right]\mod{2}.
\end{equation}
The notation EBZ stands for Effective Brillouin Zone,~\cite{moore&balents-2006} which describes one half of the Brillouin zone together with appropriate boundary conditions.  Since the BZ is a torus, the EBZ can be viewed as a cylinder, and its boundary $\partial(\text{EBZ})$ as two circles, as in Fig.~\ref{figebz}(b).  While $\mathcal{F}$ is gauge-invariant, $\bm{\mathcal{A}}$ is not, and different (time-reversal-invariant) gauges can change the sum of boundary integral by an even amount.  Heuristically, (\ref{Dk}) states that integrating the field strength of a Kramers pair of bands over half the Brillouin zone gives a topological invariant in the same way as integrating the field strength of a single band over the whole zone.  However, a gauge-dependent boundary term must be added to make the integral an integer.  As mentioned, this boundary term is ambiguous (via gauge changes) up to an even integer, so only the parity $D$ mod 2 is well-defined.

A direct proof of the existence and $\mathbb{Z}_2$ nature of the topological invariant for multiple bands without introducing gauge-dependent quantities can be obtained~\cite{moore&balents-2006} by considering maps, from the EBZ to the space of Bloch Hamiltonians, that are consistent with time-reversal, following work on the IQHE by Avron \emph{et al.}~\cite{avron&alii-1983}  The Berry field strength can be written in terms of the (gauge-invariant) projection operator onto the band pair rather than the (gauge-dependent) wavefunctions.  In this approach, the ambiguity by an even integer that is crucial to obtain a $\mathbb{Z}_2$ rather than $\mathbb{Z}$ invariant corresponds to the many different ways in which the circles that form EBZ boundaries in Fig.~\ref{figebz}(b) can be contracted to make the EBZ into a sphere.  The boundary integrals in (\ref{Dk}) just calculate the contribution to the Chern number from these ``contractions.''  On this sphere, Chern integers are well-defined for each nondegenerate band pair, but the different ways of contracting the boundaries cause the resulting integers to differ by even numbers.  An explicit numerical implementation of the Fu-Kane formula (\ref{Dk}) was given by Fukui and Hatsugai~\cite{fukuihatsugai} and will be reviewed in Section III.  

The invariant $D$ classifies the topology of band pairs; each pair carries a value of $D = 0$ or 1.  The model of graphene we study has four bands (two sites per unit cell, two spin states per site), so two band pairs $D_1$ and $D_2$, and the total band structure must have $D = 0 \mod 2$,\cite{avron&alii-1983,moore&balents-2006} so that $D_1 = D_2$.  Hence there are two phases, characterized by the value of $D = D_1$, say.  Now, if at some value of parameters the upper and lower band pairs meet somewhere in the BZ, $D$ is not well-defined, so there can be a phase transition. The two pairs with $D=1$ annihilate into two $D=0$ pairs when the bands collide as the parameters are varied.  Note that, just as total Chern number is conserved when bands collide in the IQHE, total $\mathbb{Z}_2$ is conserved here.

At half filling, these phases are both insulating, since there is a gap between the second and third bands; the semimetal that appears when the gap closes marks the phase boundary.  What happens when  disorder is added?  We now explain how the definition of the topological insulator generalizes to disordered systems.  Just as TKNN integers of a band structure give rise to Chern integers in a disordered system of finite size, the $\mathbb{Z}_2$ invariants of band pairs become ``Chern parities''.

\subsection{The topological insulator phase for Slater determinants via Chern parity}\label{supercelldef}

The TKNN integers generalize in the presence of disorder and interactions to the Chern number of the many-particle wavefunction.\cite{niu&alii-1985}  A natural question is how disorder and interactions modify the $\mathbb{Z}_2$ invariants of band pairs in spin-orbit-coupled 2D band structures.  All derivations of the $\mathbb{Z}_2$ invariants of clean systems depend on Fermi statistics in some way: for example, the existence of Kramers degeneracies and the related fact that the time-reversal operator squares to $-1$ both depend on Fermi statistics.  A many-fermion wavefunction describing an even number of fermions does not behave in the same way as single-fermion wavefunctions under time-reversal.  Hence, given only the many-fermion wavefunction, it does not seem likely that there is a generalization of the $\mathbb{Z}_2$ invariant.

However, for the particular case of many-fermion wavefunctions that are single Slater determinants of single-particle wavefunctions, the invariant can be generalized as we now show.  While the assumption of a single Slater determinant limits the treatment of interactions to the Hartree-Fock level, there is no requirement that the wavefunctions in the Slater determinant be Bloch states.  As a result, the topological insulator and QSHE can be defined and studied for any disorder strength.    

Niu \emph{et al.}\cite{niu&alii-1985} and Avron and Seiler\cite{avron&seiler-1985} showed that for disordered quantum Hall systems there exists a generalization of the TKNN invariant defined for clean systems.\cite{thouless&alii-1982}  They introduce generalized periodic boundary conditions and find an invariant Chern number, similar in form to the TKNN invariant, on the space of boundary phases.  Consider a finite system of noninteracting electrons with boundary conditions that are periodic up to phases $\phi_x,\phi_y$, as shown in Fig.~\ref{honeycomb}: this is equivalent to putting magnetic fluxes $\Phi_{x,y} = \phi_{x,y} \Phi_0/2\pi$ through the two noncontractible circles on the torus ($\Phi_0 = hc/e$ is the magnetic flux quantum).  As motivation, think of the finite system as a (possibly very large) unit cell of a lattice system.  Then in order to determine the phase of this lattice system, instead of integrating over $k$ to do the integrals in the Fu-Kane formula (\ref{Dk}), our strategy will be to integrate over the boundary phases, which introduce offsets to the wave vectors.  We first carry out this procedure, show that it reproduces the band-structure result for clean systems, and then discuss a physical picture and its relation to previous definitions.

Consider the single-particle wavefunctions of a lattice Hamiltonian such as the graphene model on a finite lattice of size $L_x \times L_y$ (see Fig.~\ref{honeycomb}).  Instead of the physical boundary conditions $\psi(\bm{x} + \bm{L}_{x,y}) = \psi(\bm{x})$ for a single-particle wavefunction $\psi$, introduce the boundary phases, or ``twists'', $\bm{\phi} = (\phi_x,\phi_y)$ via
\begin{equation}\label{bdyconds}
	\psi(\bm{x} + \bm{L}_x) = e^{i \phi_x} \psi(\bm{x}), \; 
		\psi(\bm{x} + \bm{L}_y) = e^{i \phi_y} \psi(\bm{x}).
\end{equation}
In the special case of a clean system, this shifts each allowed value of wave vector $\bm{k}$ by $\bm{\Delta} = (\phi_x/L_x,\phi_y/L_y)$.  Because it is simpler numerically to work with single-valued wavefunctions than the multiple-valued ones of equation \eqref{bdyconds}, and because the standard formalism of the Berry connection assumes the parameters to be explicit in the form of the Hamiltonian rather than living in the boundary conditions (i.e., the definition of the Hilbert space), we want to transfer the twist angles to the Hamiltonian.  This is done by making a unitary transformation to
\begin{equation}\label{unitarytransf}
	\chi(\bm{x}) = e^{-i (\phi_x x/L_x + \phi_y y/L_y)} \psi(\bm{x}).
\end{equation}
Note that for the special case $\phi_{x,y} = 2\pi$ the complex phase in \eqref{unitarytransf} is single-valued as we translate the coordinates around the torus, so that this is just a gauge transformation; if we take the phases in the square $[-\pi,\pi] \times [-\pi,\pi]$, opposite boundaries are identified under smooth gauge transformations and the twist space is a torus.  Under the change of basis \eqref{unitarytransf}, Kane and Mele's model Hamiltonian\cite{kane&mele1-2005} $H=H_0 +H_{SO} + H_R$ (see Section~I) becomes (suppressing spinor indices)
\begin{align}
	H \rightarrow \mathcal{H}(\bm{\phi}) = 
		& \sum_{\langle ij\rangle} c_{i}\!^{\dagger} \left[ t 
				+ i\lambda_R ( \bm{s}\times\hat{\bm{d}}_{ij} )_z\right]c_j 
						e^{-i \bm{\Delta} \cdot \bm{d}_{ij}} \notag \\
			& + i\lambda_{SO} \sum_{\langle\langle ij\rangle\rangle} \nu_{ij}
				c_{i}\!^{\dagger} s^z c_j
					e^{-i \bm{\Delta} \cdot \bm{d}_{ij}} \notag \\
			& + \sum_{i} (\lambda_v \xi_i + w_i) c_{i}\!^{\dagger} c_i,
\end{align}
where again $\bm{\Delta} = (\phi_x/L_x,\phi_y/L_y)$, $\bm{d}_{ij}$ is still the vector $i \rightarrow j$, and we have added a random term for on-site disorder, $w_i$, drawn from the Gaussian distribution of zero mean and standard deviation $\sigma_w$.  It is now clear that under time reversal,
\begin{equation}
	\Theta \mathcal{H}(-\bm{\phi}) \Theta^{-1} = \mathcal{H}(\bm{\phi}),
\end{equation}
since the extra phase factors in $\mathcal{H}$ change sign under complex conjugation.\footnote{Note that this means $\mathcal{H}$ is not generically time-reversal invariant.  Indeed, there are only four boundary conditions that respect time-reversal, namely those for which $\psi$ picks up a real phase upon translation around each cycle of the honeycomb lattice.  These four correspond to the TRIM of the clean system in the calculation of $D$.}  Since this directly parallels equation \eqref{trev}, the $\mathbb{Z}_2$ invariant $D$ of the Brillouin zone passes directly to twist space:
\begin{equation}\label{Dphi}
	D_\phi = \frac{1}{2\pi} 
		\left[ \oint_{\partial(ETZ)} d\bm{\phi}\cdot\!\bm{\mathcal{A}}
					- \int_{ETZ} d^2\!\phi\,\mathcal{F} \right]\mod{2},
\end{equation}
with ETZ for Effective Twist Zone, i.e., ETZ $= \{\bm{\phi}\,|\, 0\leq\phi_x\leq\pi,\, -\pi<\phi_y\leq \pi \}$.  Note that there is an independent Chern parity $D_\phi$ for each Kramers-degenerate band pair separated from the rest of the spectrum by a gap at all $\bm{\phi}$: for such an isolated pair, $\bm{\mathcal{A}}$ and ${\mathcal F}$ are defined as in \eqref{A} and after, with $\bm{k} \rightarrow \bm{\phi}$ and $u \rightarrow \chi$.

In order to make contact with the band-structure definition, we note that if there is no disorder in the Hamiltonian (i.e., $\sigma_w = 0$), there are discrete translational symmetries within the $L_x \times L_y$ supercell that induce additional non-Kramers degeneracies at some points in twist space.  With such degeneracies only the total Chern parity of all the degenerate states is well-defined.  For example, in the clean graphene model, there are only two separated sets of states, even though the dimension of the Bloch Hamiltonian increases as $L_x$ and $L_y$ increase.  Each of these sets has its own Chern parity, consistent with the counting of $\mathbb{Z}_2$ invariants in the band structure (note that the two invariants are not independent because of a zero sum rule).\cite{moore&balents-2006}  As in the band-structure case, the phase of a physical system is determined by the sum of Chern parities for occupied band pairs.  
Disorder, as discussed in the following section, breaks all degeneracies resulting from translational invariance, leaving only separated band pairs, each of which has its own Chern parity.
We now discuss to what extent Chern parities can be connected to observable quantities in a finite system.


\subsection{Charge pumping cycles in time-reversal-invariant systems}

The above definition was motivated by thinking of a finite system as a (possibly large) unit cell of an infinite periodic lattice.  The finite system was defined to be in the topological insulator phase according to whether the infinite lattice is in the topological insulator phase.  Since the original identification of this phase was by numerical observation of stable edge states in finite systems, it seems clear that there should be a direct way to detect the topological insulator in the finite system.

The total Chern number in a finite IQHE system can be interpreted as measuring the number of charges pumped when the flux through one noncontractible circle on the torus increases adiabatically by one flux quantum.  Briefly, one of the boundary phases corresponds to this driving flux, and the average over the other can be shown to yield the pumped charge.\cite{niu&thouless-1987}  The idea of ``$\mathbb{Z}_2$ pumping'' suggested by Fu and Kane~\cite{fu&kane1-2006} is the following: in a finite cylinder with boundaries, the operation of increasing the phase $\phi_x$ (in the periodic direction, around the cylinder) from $0$ to $\pi$, corresponding to a magnetic flux of one-half flux quantum through the cylinder, has the following effect in the topological insulator.  The values $\phi_x=0$ and $\phi_x = \pi$ are special because, unlike general values, they are consistent with time-reversal invariance.  At these special fluxes there are gapless states at the Fermi level that are localized near the edges because of the bulk gap.  Fermi statistics requires that these states lie in Kramers doublets.  If at zero flux, the Kramers doublet at one edge is partially occupied (has one state occupied), then the operation of changing the flux changes its occupancy to either double or zero occupancy.  Since total charge is conserved, this requires a flow of $\mathbb{Z}_2$ from one boundary to the other.  Note that in addition to the average over applied flux, there is effectively an average over the spatial extent of the edge, because the edge states are extended.  

We now give an alternate definition of the topological insulator in terms of {\it cyclic} pumping of ordinary charge.  This definition is mathematically equivalent to the definition of \ref{supercelldef} based on treating the finite system as a supercell.  In order to describe a closed pumping cycle, we need to add a second stage to the Fu-Kane process of increasing boundary phase $\phi_x$ from $0$ to $\pi$: although both these phases are consistent with time-reversal invariance, physical properties, like the occupancy of an edge doublet, are not identical at these different values of $\phi_x$ (even at the same $\phi_y$).  The only requirement on the second stage is essentially that it return the system to its original state {\it without applying a time-reversal-breaking flux}.
The second stage gives a closed pumping cycle that returns the system to its original Hamiltonian, which means that the amount of charge pumped in a specified cycle is a well-defined integer quantity.

\begin{figure}[!ht]
	\centering
	\scalebox{0.38}{\includegraphics{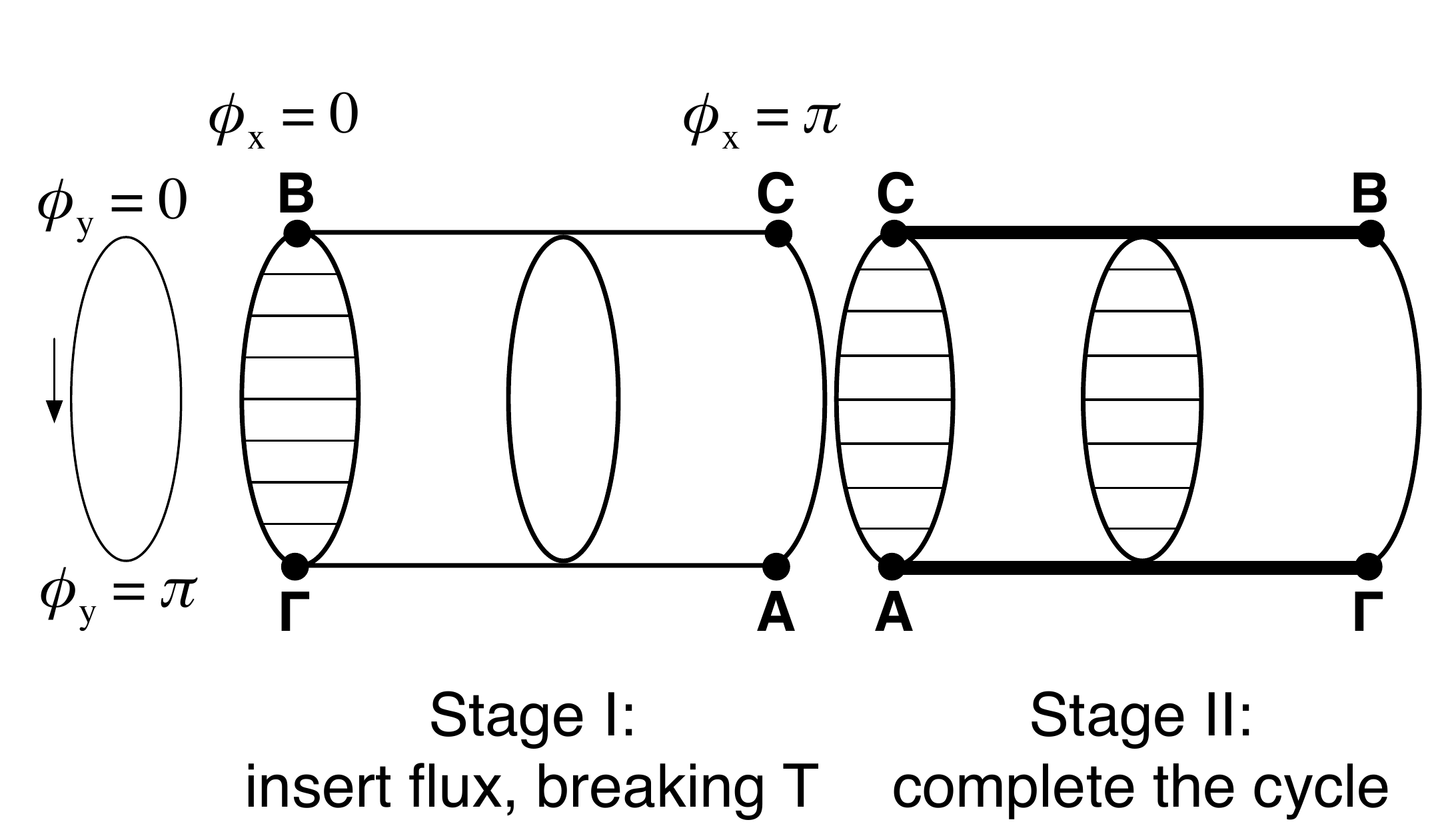}}
	\caption{Graphical representation of charge pumping cycle for Chern parities.
		The first stage takes place on the ETZ (as in equation \eqref{Dphi}), and
		 the flux $\phi_x$ increases adiabatically from 0 to $\pi$.
		In the second stage the Hamiltonian at $(\phi_x=\pi,\phi_y)$ is adiabatically transported through the space of Hamiltonians to return to the Hamiltonian at $(\phi_x=0,\phi_y)$.  The
		difference between the second stage and the first is that at every step of the second stage, the Hamiltonians obey the time-reversal conditions required at $\phi_x=0$ or $\phi_x=\pi$.  The bold lines indicate paths along which all Hamiltonians are
time-reversal invariant, and the disk with horizontal lines indicates, as
before, how pairs of points in the second stage are related by time-reversal.
		\label{figclosed}}
\end{figure}

Although the number of charges pumped is dependent not only on the first stage but on the second stage as well, whether this is an even or odd number is entirely determined by the first stage, as we now show.  A closed pumping cycle is shown in Fig.~\ref{figclosed}.  (As before, we discuss the cycle in terms of variable Hamiltonians in a fixed Hilbert space rather than in terms of changing the boundary conditions on Hilbert space.)  The original physical system's ETZ is the first stage of the cycle: the Hamiltonians are functions of $\phi_x$ from 0 to $\pi$ and $\phi_y$ from $-\pi$ to $\pi$, with time-reversal constraints that act on the boundary circles at $\phi_x = 0$ and $\phi_x = \pi$.  If $\phi_x$ takes one of these values, then the Hamiltonian at $\phi_y$ is time-reversal conjugate to that at $\phi_y^\prime = - \phi_y$.  The second stage can be any continuous change of the Hamiltonians that takes the $\phi_x=\pi$ system back to the $\phi_x = 0$ system and always satisfies the conjugacy condition between $\phi_y$ and $\phi_y^\prime$.  This is the key difference between the second stage and the first stage: for intermediate values $0 < \phi_x < \pi$, there is no such conjugacy condition.  The physical interpretation is that the second stage should be possible without introducing flux through the first noncontractible circle.

Now the torus shown in Fig.~\ref{figclosed} has one Chern integer for each isolated band pair. Summing over occupied bands gives the amount of charge pumped in the cycle.  Although this integer charge depends on the second stage, its parity is solely determined by the first stage, i.e., the physical system.  In particular, for the ordinary insulator there is some closed cycle that pumps zero charge, while for the topological insulator there is some closed cycle that pumps unit charge.  These results follow from the same proof as for the band structure case in Ref.~\onlinecite{moore&balents-2006}: one shows that the differences in resulting Chern integers between any two second stages are even.  The pumping definition gives some physical intuition for the ``contractions'' introduced there; instead of contracting the EBZ to a sphere, here the ETZ is contracted to a torus by adding the second stage.  The technical reason that these two constructions are equivalent is that, since the appropriate spaces of Hamiltonians are contractible (i.e., have $\pi_1 = 0$), the two closed manifolds, the torus and the sphere, both have the same topological invariants, namely integer-valued Chern numbers.

The topological insulator in disordered systems has been studied previously by locating the transition between topological and ordinary insulators as a point or region where the localization length of single-electron eigenstates diverges.  The existence of the topological insulator is inferred from the existence of this transition region (or alternately from the edge states in the topological insulator phase).  In principle this approach is different from ours, in that our definition probes the existence not just of extended states but specifically of extended states that contribute to the pumping of charge, or alternately can be expected to give rise to edge states.  The same distinction arises in the quantum Hall effect, where looking for extended states of nonzero Chern number is a more direct probe of quantum Hall physics than considering the inverse participation ratio, for example, which would detect extended states of zero Chern number in addition to topological states.  However, as in the quantum Hall case, we find that the phase boundaries from our Chern-parity definition are consistent with those obtained from calculations of the localization length.~\cite{onoda&alii-2006,obuse&alii-2007}
\begin{figure*}[t]
	\scalebox{0.4}{\includegraphics{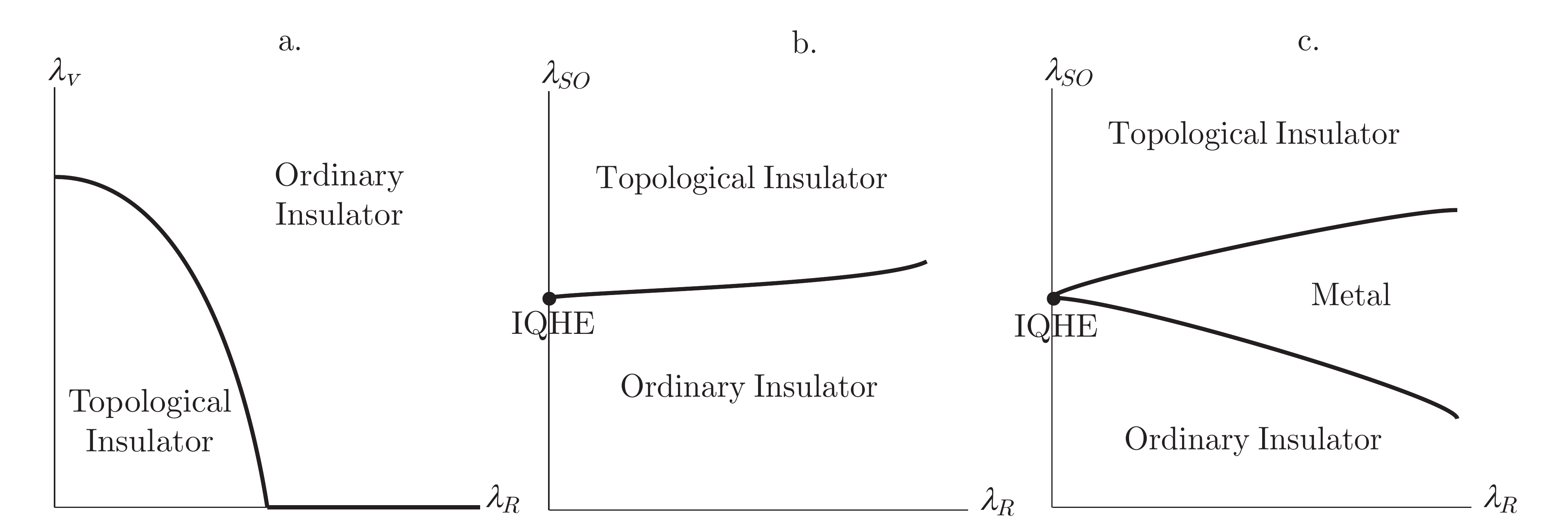}}
	\caption{Schematic phase diagram.  (a) Phase diagram for a clean system, with
	fixed
	$t$ and $\lambda_{SO} \neq 0$ (after Kane and Mele\cite{kane&mele2-2005}).
	(b) Again a clean system, now with fixed $t$ and $\lambda_v \neq 0$.
	(c) The expected form of the phase diagram at nonzero disorder (we run all
	simulations at fixed $\lambda_v$).  The phase boundary in (b) opens up
	into a metallic phase, closing only when $\lambda_R = 0$, where there should be
	an IQHE transition.
	\label{schematic}}
\end{figure*}


\section{Graphene Model and Numerics}
\subsection{The phase diagram of the disordered graphene model}

It is useful to review some general expectations before applying the definitions of the previous section to study a disordered version of the graphene model.  In two-dimensional systems with time-reversal invariance and no spin-orbit coupling, even very weak disorder will localize electron wavefunctions, so that these systems do not ever conduct in the thermodynamic limit.  In the presence of a magnetic field, as in the IQHE, there are isolated energies with extended states, but no finite-width range of energies with extended states, and hence no true metallic phase.  Hikami \emph{et al.}~showed\cite{hikami&alii-1980} that disordered two-dimensional systems \emph{with spin-orbit coupling} can nevertheless support a metallic phase, referred to as a ``symplectic metal''.  In our case, any metallic phase would presumably appear in a region around the parameter set that closes the (clean) gap, as depicted schematically in Fig.~\ref{schematic}.

We also know that at $\lambda_R=0$ the $z$ component of spin is a good quantum number ($s^z$ commutes with $H$), so the system reduces to two copies of the Haldane model,\cite{haldane-1988} which has a quantum Hall plateau transition with no metallic phase. 

\subsection{Lattice Implementation}

For numerical work we use the algorithm of Fukui and Hatsugai, which we review here.\cite{fukuihatsugai}  The formula \eqref{Dk} for $D$ requires a gauge choice for the Hamiltonian eigenstates at each $\bm{\phi}$ on the two boundaries of the half-torus $0\leq\phi_x\leq\pi,\, -\pi<\phi_y\leq \pi$.  That is, the ``field strength'' $\mathcal{F}$ is gauge invariant, but the gauge potential $\bm{\mathcal{A}}$ is not.  The eigenstates form Kramers pairs related by time reversal, and the gauge choice must respect this constraint.
\begin{figure}[!ht]
	\scalebox{0.4}{\includegraphics{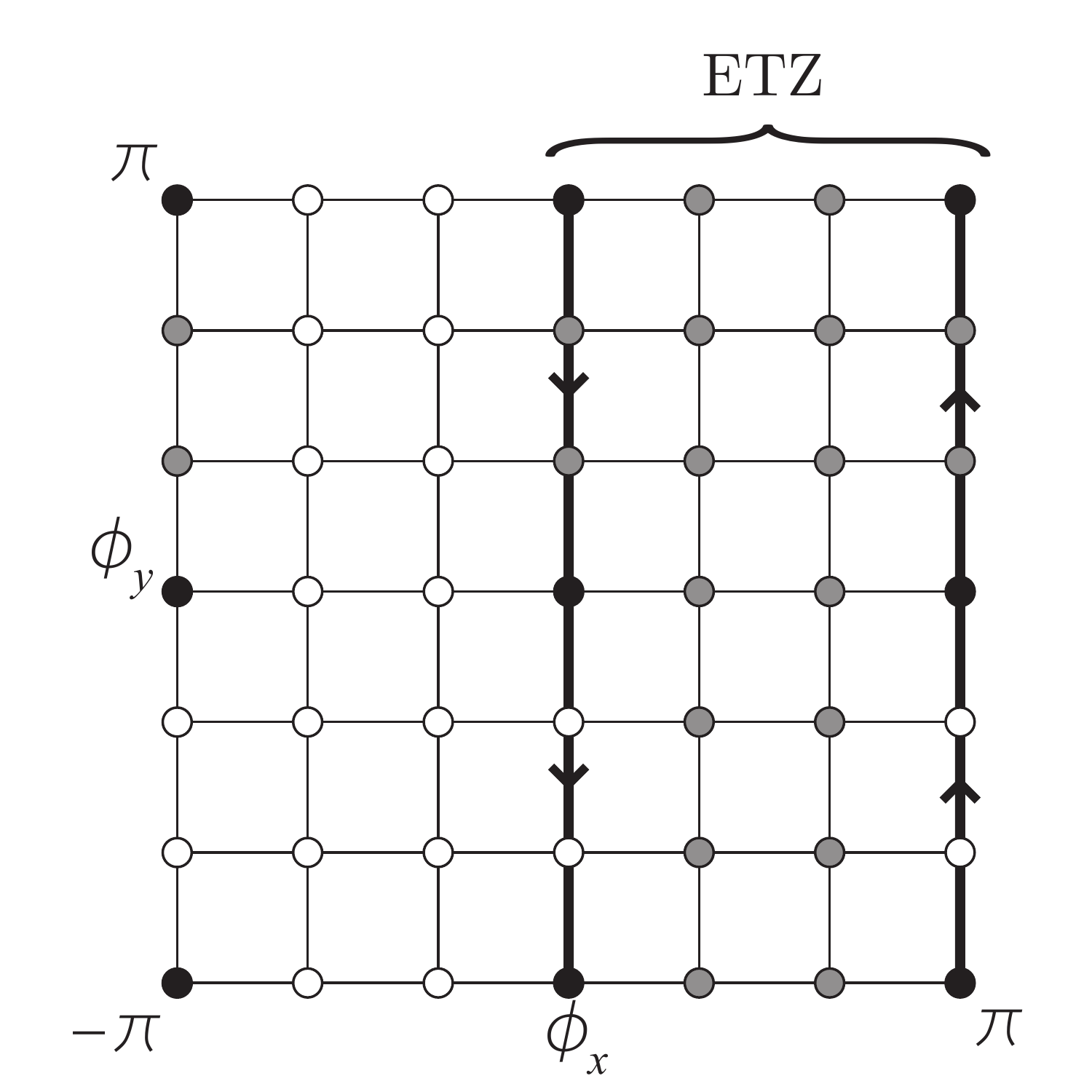}}
	\caption{Twist Space.  The bold lines indicate the boundaries of the ``Effective
	Twist Zone'', the region we integrate (or sum) over to calculate the 
	Chern parity.  The arrows indicate the direction to perform the sum
	over the boundary terms, and the lattice sites in gray indicate those for which
	the Hamiltonian eigenvectors are independently specified.  That is, time reversal
	symmetry determines the eigenvectors on the white sites once those at the gray
	sites are found.
	\label{twistspace}}
\end{figure}

Now, at the time-reversal invariant points $\bm{\phi} = (0,0)$, $(0,\pi)$, $(\pi,0)$, and $(\pi,\pi)$, the solid points in Fig.~\ref{twistspace}, the spectrum is degenerate, with two states at each energy. The gauge condition requires that $\Theta$ interchange the two with a phase factor $e^{\pm i\pi/2}$ (so that $\Theta^2 = -1$ as required for single-fermion states).  Numerical diagonalization will not, in general, return eigenvectors that obey this condition, but we can force them to do so as follows: choosing one of the two members of each Kramers pair at energy $\varepsilon_{2n-1}=\varepsilon_{2n}$ and calling that vector $\chi_{2n-1}$, we discard the other and replace it by
\begin{equation}
	\chi_{2n} = \Theta \chi_{2n-1}.
\end{equation}
On the rest of the boundary, eigenvectors can be chosen freely on $0<\phi_y<\pi$.  On $-\pi<\phi_y<0$ the algorithm takes
\begin{equation}
	\chi_{n}(-\bm{\phi}) = \Theta \chi_{n}(\bm{\phi}).
\end{equation}
In summary, the algorithm leaves alone the results of numerical diagonalization at all the gray points in Fig.~\ref{twistspace}, and by hand enforces the gauge condition on the rest of $\partial(\text{ETZ})$.\footnote{The points $(\phi_x,\pi)$ and $(\phi_x,-\pi)$ are not physically distinct; as pointed out earlier, a gauge transformation relates $\mathcal{H}(\phi_x,-\pi)$ to $\mathcal{H}(\phi_x,\pi)$ as given, for example.  We therefore impose $\mathcal{H}(\phi_x,-\pi) \equiv \mathcal{H}(\phi_x,\pi)$ in the calculation.  Alternatively, we could introduce boundary terms along $\phi_y = \pi,\,-\pi$ to compensate for the discrepancy.}

With the eigenstates fixed at each point on the ETZ, we follow Fukui and Hatsugai and construct $U(1)$ parallel transporters on the links, as
\begin{equation}
	U_x(\bm{\phi}) = \frac{g_x}{|g_x|},\;
		g_x = \det{\bm{\chi}^\dagger(\bm{\phi})\bm{\chi}(\bm{\phi}+\hat{\bm{x}})}
\end{equation}
and $U_y$ similarly.  Like $\bm{u}$ in \eqref{A}, $\bm{\chi}$ is a matrix built from occupied state vectors, and $\hat{\bm{x}}$ translates by one link in the $\phi_x$ direction.  In the continuum limit $g$ should approach a pure phase, but for non-zero lattice spacing it will in general have $|g|\leq 1$, since the occupied subspace of interest will not embed in the total Hilbert space in the same way at every lattice point.

In the end, the only retained information will be the variation of the relative phases (hence the definition of $U$), which can be captured by choosing a lattice constant so small that the phase field varies slowly over one link.  However, the scale of variation will presumably differ for different disorder realizations, and we would like a way to diagnose this and throw out those realizations for which fast variation makes the calculation unreliable.  The phase is periodic, and if it were to wind through $2\pi$ over the distance of one link the algorithm would miss this fact, so the relative phase will not provide a good diagnostic.  As a proxy, we choose to cull out disorder realizations that result in small determinants in most simulations, since a small overlap between adjacent occupied eigenspaces indicates rapid variation.  Of course, this filtering could introduce a selection bias into the results; these effects are within the statistical uncertainty of our analysis (in particular, within the error bars of Fig.~\ref{widthscaling}).

Associated with the transporter on each link is a gauge potential $A_{x,y} = \log U_{x,y}$.  This $A$ is pure imaginary, and the logarithm is defined to return the branch $A/i \in (-\pi,\pi)$.  Associated with the transport around each plaquette is a flux
\begin{equation}
	F(\bm{\phi}) = \log U_x(\bm{\phi}) U_y(\bm{\phi}+\hat{\bm{x}})
		U_{x}^{-1}(\bm{\phi}+\hat{\bm{y}}) U_{y}^{-1}(\bm{\phi}),
\end{equation}
again satisfying $F/i \in (-\pi,\pi)$.  With these definitions in hand, the lattice $\mathbb{Z}_2$ invariant corresponding to $D_\phi$ \eqref{Dphi} is
\begin{equation}\label{DL}
	D_L = \frac{1}{2\pi i} \left[ \sum_{| \in \partial(ETZ)} A_y - 
		\sum_{\square \in ETZ} F \right] \mod 2.
\end{equation}
Of course, the sum over the boundary should have the same orientation as the corresponding contour integral in \eqref{Dphi}; in Fig.~\ref{twistspace}, this means the sum on the left boundary should carry a minus sign, following the arrows.  As mentioned previously, this formalism has the desirable property of guaranteeing that $D_L = 0 \:\mathrm{or}\: 1$, but using too coarse a mesh can return the wrong value.

\subsection{Numerical Results}

\begin{figure}[!ht]
	\scalebox{0.3}{\includegraphics{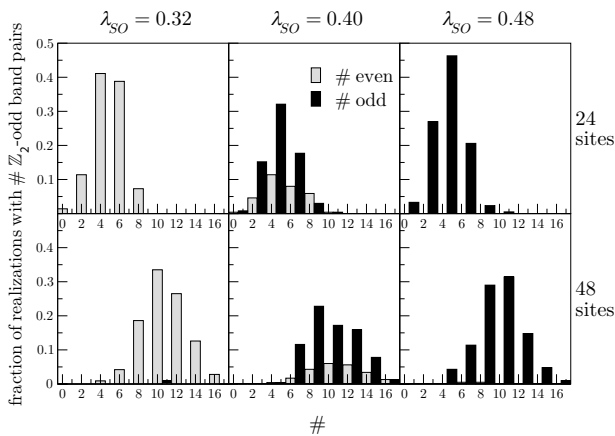}}
	\caption{Distribution of Chern parities for band pairs.
		The bar heights represent the fraction of disorder realizations (out of 
		$\sim 200$ trials) that
		have a given number $\#$ of band pairs with $D_L = 1$ in the occupied
		(half-filled) subspace.  Those with an even number $\#$ will have an overall 
		$D_L = 0$, those with an odd number will have overall $D_L = 1$.  All
		these simulations were done with $t = -1$, $\lambda_R = \lambda_v = 1$,
		and $\sigma_w = 0.3$.  Reading across, $\lambda_{SO}$ increases and the system
		transitions
		from all realizations having even parity at small $\lambda_{SO}$ to odd parity
		 at large $\lambda_{SO}$.  Reading down, doubling the system size
		doubles the total number of Kramers pairs and roughly doubles the number of 
		$\mathbb{Z}_2$-odd pairs.
		\label{histograms}}
\end{figure}

As noted after equation \eqref{Dphi}, the number of nondegenerate Kramers pairs in a disordered system will generically be extensive, and we can use equation \eqref{DL} to calculate the Chern parity of each pair separately.  Fig.~\ref{histograms} shows the results of such calculations on a $4\times 6$ lattice to present a picture of the three phases: normal insulator, symplectic metal, and topological insulator.  In all phases, there are Kramers pairs with $D_\phi = 1$ in the lower half of the energy spectrum, as indicated by the bars.  However, in the normal insulator ($\lambda_{SO}$ well below the transition, or gap closure) there are an \emph{even} number of such pairs in the occupied half of all realizations, so that the overall Chern parity is even.  The presence of a very small number of realizations (2 of 215 for a $6\times 8$ lattice) with odd parity indicates either the tail of the disorder-broadened transition or, more likely, that the weak filter applied for these simulations failed to catch all realizations with rapidly varying link variables.

When $\lambda_{SO}$ is large, almost all realizations have an odd number of $\mathbb{Z}_2$-odd Kramers pairs (as with the small-$\lambda_{SO}$ case, 2 realizations do not follow the rule), and in the region near the transition of the clean system there are instances of both types.  The presence of disorder causes the ``gap'' to close at different values of $\lambda_{SO}$ for different realizations, and also at different energies.  The latter fact means that extended states are present throughout a finite spread of energies, while the former means that the metallic state exists over a finite region of parameter space.

For an integer quantum Hall system, Yang and Bhatt\cite{yang&bhatt-1996} have shown how to extract the localization length exponent $\nu$ from such calculations, in their case the Chern numbers of Landau sublevels.  Specifically, sublevels with non-zero Chern number contain extended states, which should only occur at isolated energies in the IQHE.  Therefore, the number $N_c$ of such sublevels should decrease as the system size $N_s$ increases, and in fact $\langle N_c \rangle \propto N_s^{1-1/2\nu}$, where $\langle \rangle$ indicates an average over disorder realizations.  A similar approach for the QSH system here should reveal $\langle N_D \rangle \propto N_s$, where $N_D$ is the number of $\mathbb{Z}_2$-odd Kramers pairs, since we expect a stable metallic band of energies in the thermodynamic limit (as observed by Obuse \etal\cite{obuse&alii-2007} and Onoda \etal\cite{onoda&alii-2006}).  With enough data and large enough systems, finite-size-induced broadening of the edges of this band should also make $\nu$ accessible via a subleading term in the scaling.  We find that larger system sizes require a finer mesh in twist space for these Kramers-pair-resolved simulations to return stable results, so that the requirements quickly outstrip our resources.  Nevertheless, comparison of the two rows in Fig.~\ref{histograms} indicates that the location of the mean roughly doubles.  This is what would be expected for the middle case given the above considerations;  the mean for the topological insulator (the right-hand panels in Fig.~\ref{histograms}) should grow more slowly, as in the case studied by Yang and Bhatt.

The total phase of the system, given by the total Chern parity, is more relevant to possible measurements than the Chern parity of each Kramers pair.  The former maintains its meaning if we consider the ground state wave function of the many-electron system, formed as a Slater determinant of the single-electron states we use here.  There is also a computational benefit to calculating $D_\phi$ for the whole occupied subspace rather than for individual pairs, as well --- at larger system sizes, and also at stronger disorder, the link variables for the half-filled subspace vary much more slowly than those for the individual Kramers pairs, making the calculation more robust.  For these reasons the remaining plots in this paper depict only the Chern parity of the half-filled system.

\begin{figure}[!ht]
	\scalebox{0.4}{\includegraphics{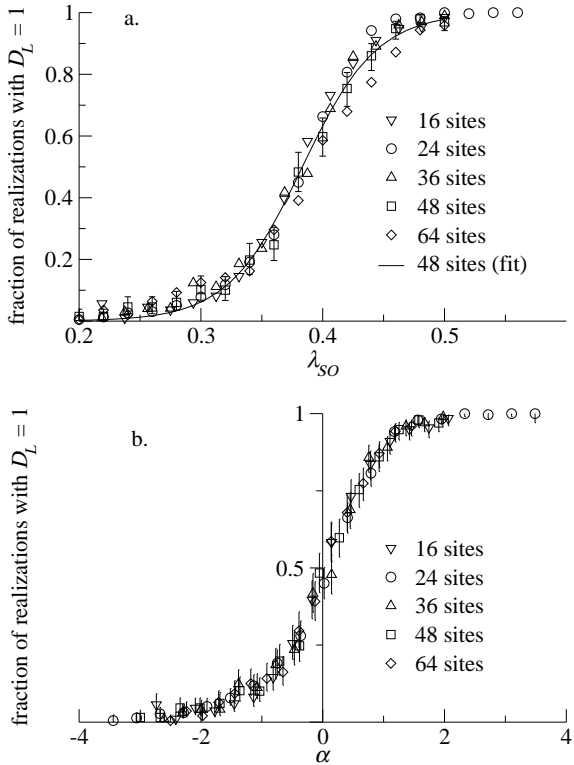}}
	\caption{(a)  At finite $\lambda_R$, the ``metallic'' region persists as the
	 system size grows, and even broadens in the case shown here ($t = -1$, 
	 $\lambda_R =\lambda_v =\sigma_w=1$).  We identify the metallic
	 region as those values of $\lambda_{SO}$ for which some, but
	 not all, disorder realizations have $D_L = 1$.  As explained in the text,
	 the fit to the simulation data (shown only for $6\times 8$ systems here) has 
	 the form $(\tanh \alpha + 1)/2$, with 
	 $\alpha = m(\lambda_{SO} - \lambda_{SO}^*)$.
	 The error bars represent 95\% confidence intervals assuming a binomial
	 distribution of outcomes for each $\lambda_{SO}$.
	 (b)  The scaling collapse of the data in (a), based on the best fit $m$ and
	 $\lambda_{SO}^*)$ for each system size.
	 \label{stablemetal}}
\end{figure}

To show that a metallic region of non-zero extent in parameter space exists in the thermodynamic limit, we need to verify that the mixed-phase region does not shrink to zero as we increase the system size.  Figure~\ref{stablemetal}(a) shows that as we make the system larger, the transition region certainly does not get narrower, and in fact the largest system size seems to have the broadest transition.

\begin{figure}[!ht]
	\scalebox{0.4}{\includegraphics{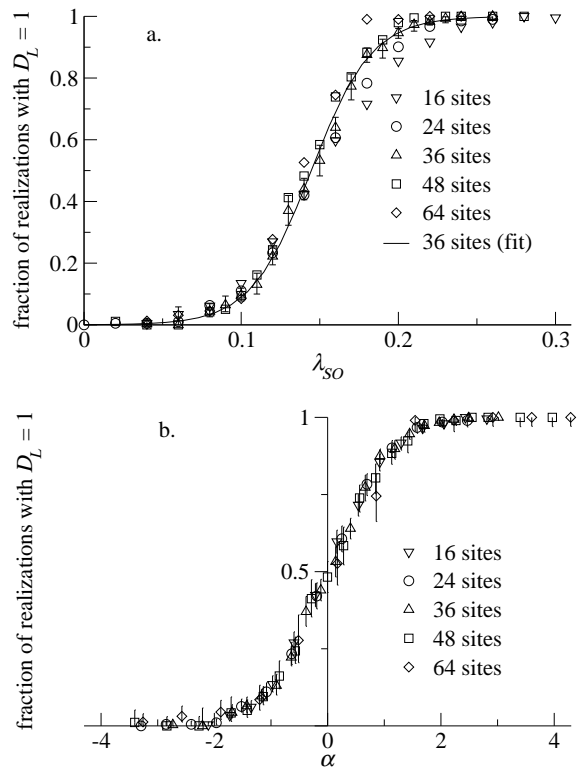}}
	\caption{(a) At $\lambda_R = 0$, the metallic region gets narrower as the system 
		size increases.  The fit is to a $6\times 6$ system and $\lambda_R = 0$; as in
		Fig.~\ref{stablemetal}, $t = -1$, $\lambda_v = \sigma_w = 1$.
		(b)  The scaling variable $\alpha$ again comes from fitting to a tanh, 
		and error bars represent 95\% confidence intervals.
		\label{haldanemodel}}
\end{figure}

We can quantify the scaling of the metallic region's width with system size by assuming a simple one-parameter scaling form for the curves in Fig.~\ref{stablemetal}(a) and defining the width of the curve to be proportional to the reciprocal of the maximum slope: width $\sim 1/$slope.  With sufficient data, one could expand the approximately linear region near the middle of the transition in a power series with a few coefficients as fit parameters.  Since our simulation data are limited, we opt instead to assume the form $(\tanh \alpha(s) + 1)/2$, which has roughly the right shape.  If $\alpha = m(\lambda_{SO}-\lambda_{SO}^*)$, then $m$ is exactly the maximum slope we want.  Figure~\ref{stablemetal}(b) plots the data versus the best-fit scaling variable $\alpha$ for each system size; the points appear to fill out a smooth curve, justifying the scaling hypothesis.  The best fit for $m$ and $\lambda_{SO}^*$ is determined by minimizing a weighted $\chi^2$ statistic.\cite{PDGstatistics-2006}  In particular, we assume that the results ($D_L = \pm 1$) of independent simulations at a fixed parameter set are distributed binomially, and that for each system size there are the same number of trials at each value of $\lambda_{SO}$ (which is roughly true).  In that case, the variance of the distribution can be estimated as $\sigma^2 \propto p(1-p)$,\cite{d'agostini-2003} where $p$ is the fraction of disorder realizations returning $D_L = 1$, and then $1/\sigma^2$ is an appropriate weight for the statistical test.  The error bars in Figs.~\ref{stablemetal} and \ref{haldanemodel} are also assigned based on a binomial model of the data.\cite{PDGstatistics-2006}

By contrast, at $\lambda_R=0$ the Hamiltonian \eqref{ham} reduces to two copies of the Haldane model and so should exhibit the quantum Hall plateau transition, which  looks like a step function at zero temperature in the thermodynamic limit.  In Fig.~\ref{haldanemodel}(a) the width of the transition region shrinks as the system size grows, consistent with the prediction.

More quantitatively, Pruisken\cite{pruisken-1988} has shown in a renormalization group framework that the functional form of the crossover for the IQHE looks like 
\begin{equation}
	p(L,B) = f(\alpha), \quad \alpha \propto L^{1/\nu} (B-B^*),
\end{equation}
where $L$ is the linear size of the finite sample, $B$ is the applied magnetic field, and $\nu$ is again the localization length exponent.  The function $p$ could be either the longitudinal or transverse conductivity in the IQHE.  Therefore, $p(\infty,B^* \pm \varepsilon) = f(\pm \infty)$, i.e., the transition is sharp in the thermodynamic limit.  In our system the analogous parameter to $B$ is $\lambda_{SO}$: in the Haldane model, the spin-orbit coupling breaks time-reversal invariance locally, like $B$ does in the IQHE.  The width of the transition region in $B$ is governed by the way the Landau level energies respond to changing $B$, and the width of the transition region in our model (for fixed $\lambda_v$) is determined by the response of the gap to changing $\lambda_{SO}$.  Since both responses are linear, we expect that the appropriate scaling variable will be $\alpha \propto L^{1/\nu} (\lambda_{SO}-\lambda_{SO}^*)$.  

\begin{figure}[!ht]
	\scalebox{0.31}{\includegraphics{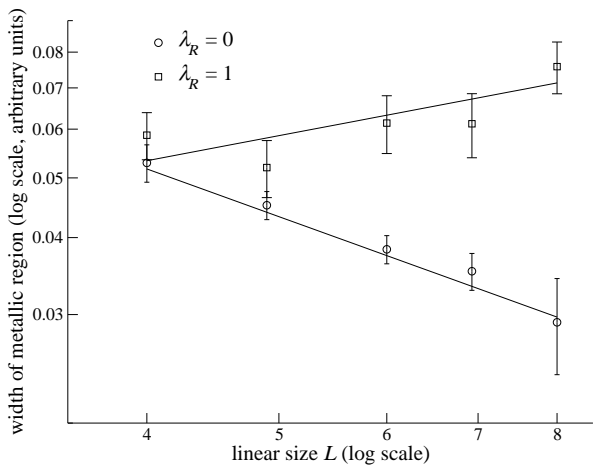}}
	\caption{Width of the metallic transition region as a function of linear size 
	$L$.  The width is determined by the reciprocal of the maximum slope of the
	simulation data in Figures \ref{stablemetal} and \ref{haldanemodel}, and the
	linear size is taken to be the square root of the number of sites.
	\label{widthscaling}}
\end{figure}

This form  would allow us to extract the exponent $\nu$ from the scaling of the maximum slope with system size for large systems (there are corrections at small system sizes).  Again making the fit to a tanh described above, Fig.~\ref{haldanemodel}(b) shows the scaling form, and Fig.~\ref{widthscaling} shows the scaling of width with linear system size.  In particular, a regression gives $1/\nu = 0.78 \pm 0.03$, to be compared with the accepted value of $1/\nu \approx 0.42$.  That is far from good agreement, but the observed scaling should not be taken as implying a new universality class.  (For reference, the network-model work by Obuse \etal\cite{obuse&alii-2007} found $1/\nu \approx 0.37$, and Onoda \etal\cite{onoda&alii-2006} recently found a value $1/\nu \approx 0.63$)  First, there should be finite-size corrections to the simple scaling assumed for the small systems considered here.  Second, the scaling form is in principle different for different geometries, and the simulations were done for systems of varying aspect ratio (1 to 1.5).  Nevertheless, it is clear that the qualitative behavior at $\lambda_R=0$ is as expected, showing no metallic phase, and the behavior at $\lambda_R=1$ is consistent with the presence of a metallic region. 

\begin{figure}[t]
	\scalebox{0.31}{\includegraphics{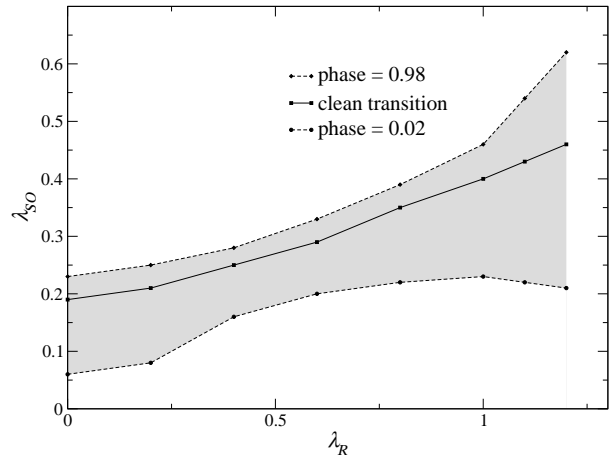}}
	\caption{Approximate width of metallic region for a 
		$4\times 6$ lattice (fixed $\lambda_v=\sigma_w=1$).  The dashed curves
		indicate the parameter values at which
		$98\%$ and $2\%$ of the disorder realizations have $D_L = 1$.  This
		underestimates the true width of the ``metallic'' region but hopefully avoids
		some amount of the inevitable error due to small system size.  This diagram
		 should
		offer a reasonable approximation to the thermodynamic (infinite-system) phase
		diagram away from $\lambda_R = 0$.  At $\lambda_R = 0$, there should be no
		 metallic phase, but 
		a sharp transition between the two insulating phases in the thermodynamic limit.
		Given the plots in Fig.~\ref{stablemetal}, it
		appears that finite size effects also reduce the width of the metallic phase
		at large $\lambda_R$.
		\label{widthplot}}
\end{figure}

Finally, by varying $\lambda_R$ and noting the $\lambda_{SO}$ values that mark the edges of the transition region for each $\lambda_R$, we can map out the phase diagram of the Hamiltonian \eqref{ham}, as in Fig.~\ref{widthplot}.  
The widths obtained this way are in rough agreement with the scaling analysis outlined above, which returns the behavior of the width and not its normalization.  As the results in Figures \ref{stablemetal} and \ref{haldanemodel} show, this phase diagram will overestimate the width of the metallic phase at $\lambda_R=0$, which is really zero.  Together, these simulations confirm the expectation of Fig.~\ref{schematic} within the accuracy of our computational methods.

\section{Summary}

Previous work\cite{kane&mele1-2005,kane&mele2-2005,roy,fu&kane1-2006,moore&balents-2006} defined a $\mathbb{Z}_2$ topological invariant in infinite lattices that is similar to the TKNN invariant for the integer quantum Hall effect.\cite{thouless&alii-1982}  In disordered systems with boundaries, Fu and Kane~\cite{fu&kane1-2006} defined a topological invariant in terms of pumping of the occupancy of Kramers-degenerate edge states.  We have given a definition of a topological invariant valid for disordered systems without boundary, i.e., without appeal to edge states.  The ``Chern parity'' can be thought of as describing either a finite system with boundary phases or an arbitrarily large supercell in an infinite lattice system with well defined wavevector.  A physical effect of Chern parity is that it determines whether the amount of charge pumped in a certain type of closed pumping cycle is even or odd.  The ``ordinary'' and ``topological'' insulator phases can be distinguished by this invariant as long as many-body effects do not prevent description of the ground state as a Slater determinant.  Chern parity in the spin quantum Hall effect is the natural generalization of Chern number in the integer quantum Hall effect.

In a disordered system, the only degeneracies expected to survive are Kramers degeneracies at time-reversal invariant values of the boundary phases.  In this case, each pair of states related by Kramers degeneracies can be assigned its own Chern parity, and the overall Chern parity of all occupied state pairs determines the observable phase.  The lattice algorithm for $\mathbb{Z}_2$ topological invariants laid out by Fukui and Hatsugai\cite{fukuihatsugai} allows numerical identification of the topological insulator phase in disordered QSH systems.  Implementing this algorithm for the specific graphene model Hamiltonian of Kane and Mele\cite{kane&mele2-2005} with added on-site disorder, we observe the ordinary and topological insulator phases in simulations.  While the number of ``odd'' pairs (state pairs with odd Chern parity) varies with the disorder realization, there is an even number of odd pairs in the ordinary insulator, and an odd number of odd pairs in the topological insulator.

We find that a metallic phase opens up between the two insulating phases for generic spin-orbit coupling.  This agrees with the prediction of Hikami \etal\cite{hikami&alii-1980} that spin-orbit coupling can protect a 2D metallic phase from disorder, and confirms the simulation results of Onoda \etal\cite{onoda&alii-2006} and Obuse \etal\cite{obuse&alii-2007}  The methods in this paper could in principle be used to study the three-dimensional case and confirm the argument in Ref.~\onlinecite{fu&kane2-2006} that only one of the four invariants of a band structure~\cite{moore&balents-2006,rroy3d,fu&alii-2007} is stable to disorder.  While there is now strong evidence~\cite{obuse&alii-2007} that the phase transitions in the 2D QSHE are, except for special points, in the previously studied symplectic metal-insulator class, there is as yet no numerical study of the phase transitions in three-dimensional topological insulators.

\acknowledgments

The authors thank L.~Balents, H.~Haggard, D.-H.~Lee, C.~L.~Kane, and A.~Vishwanath for useful comments and acknowledge support from NSF DMR-0238760 and an NDSEG fellowship (A.~E.).  Numerical work took place on a cluster donated by the IBM SUR program.

\end{document}